Bell's inequality violation due to misidentification of spatially non stationary random processes

Louis Sica

Code 5630
Naval Research Laboratory
Washington, D. C. 20375 USA
(202)-767-9466
e-mail: sica@ccs.nrl.navy.mil

**Abstract**

Correlations for the Bell gedankenexperiment are constructed using probabilities given by quantum mechanics, and nonlocal information. They satisfy Bell's inequality and exhibit spatial non stationarity in angle. Correlations for three successive local spin measurements on one particle are computed as well. These correlations also exhibit non stationarity, and satisfy the Bell inequality. In both cases, the mistaken assumption that the underlying process is wide-sense-stationary in angle results in violation of Bell's inequality. These results directly challenge the wide-spread belief that violation of Bell's inequality is a decisive test for nonlocality.

## 1. Introduction

The present article is concerned with the predictions of quantum mechanics regarding ideal Bell's inequality experiments which use 100% efficient detectors, and which have no collection loopholes. Although past experiments [1] with 90% and higher singles rates can be accounted for with local realistic models [2,3], there is a widespread belief going back to Bell [4], and based on extensive experimental agreement with quantum mechanics, that if perfect correlation experiments were performed, the results would agree with quantum mechanical predictions. Indeed, such improved experiments have been proposed [5] and are on the way to realization. Their agreement with quantum mechanics would rule out models for which singles rates are an important component.

Bell's theorem [6] was originally derived from a number of physical and statistical assumptions as a constraint on statistical correlations. The experimental context of the theorem was the Bohm version [7] of the EPR gedankenexperiment [8] schematized in Fig. 1. In this experiment, identical particles in a singlet state fly out from a source in two different directions and encounter Stern-Gerlach spin measuring apparatuses oriented at angles $\theta_A$ and $\theta_B$. Particles emerge from these spin meters in two discrete directions indicating a spin of $\pm 1$ in units of $\hbar/2$. When $\theta_A = \theta_B$, the measured spin values $A$ and $B$ are always equal and opposite. This follows from quantum mechanical properties of the singlet state, which has the same form for every pair of equal angular settings of the spin meters. However, Bell deduced that for spin-meters pointing in arbitrary directions, the correlation of output readings is given by

$$\langle A(\theta_A)B(\theta_B) \rangle = -Cos(\theta_A - \theta_B). \tag{1}$$

He was not able to create a local model for this correlation, i.e., a model based on the assumption that the spin meters act independently on particles prepared from common initial conditions at a past time. However, he found that he could account for the correlation if information was transmitted instantaneously from one spin measuring device to the other at the time of measurement. (Such information is defined as nonlocal because it travels faster than the signal velocity of light, and is not attenuated with distance.)  Bell sought to generalize this finding through the derivation of an inequality that three correlations must satisfy if measurements are based only on common local information shared by the particles when they separate.  Violation of Bell's inequality by experimentally verified correlations is then interpreted to imply that no such local model can account for (1).

However, it is now known that some of Bell's assumptions are extraneous to the derivation of the Bell inequality and that it is a completely general numerical relation that may be derived independently of probability and statistics, and must hold equally well for deterministic sequences. Eberhard [9] derived Bell's inequalities without hidden variables on the basis of counterfactual reasoning. Sica [10,11] emphasized that Bell's inequalities are limits of arithmetic inequalities that must be satisfied by any cross-correlations of three or four finite data-lists (as appropriate), with data restricted to ±1. Further, since experimental data are inherently finite, such data arranged in appropriate lists cannot violate the Bell inequality, no matter how exotic the statistics. Correlations of both deterministic and random data are thereby constrained.  Thus understood, the Bell's inequality provides a tool that may be used to uncover characteristics of quantum correlations that have previously gone unnoticed.

In [12], the author showed that in spite of the fact that (1) holds for correlations of two measurements carried out at $\theta_A$ and $\theta_B$ on opposite sides of the apparatus of Fig. (1), additional pair correlations among three or more variables cannot all be of the form (1).  Thus, the process defined by the experiment as a whole is not stationary in second order correlations, even though it appears to be on the basis of the first two measurements. (These first two measurements on two different particles commute, while any further measurements, real or counterfactual are not commutative with them.)  It is this feature of the Bell correlations, that (1) holds for two measurements suggesting stationarity in angle coordinates, but does not hold for all variable pairs when there are more than two measurements, that has made the Bell theorem difficult to understand. Some understanding of how this comes about may be obtained from examples.  Three and four variable Bell's inequalities were considered in [12], with all correlations based on measured data.

The predicted correlations were found to be nonstationary, and to satisfy Bell's inequalities. The author believes that the procedure used in [12] is as close to Bell's intent as can be achieved in actual experiments.

Nevertheless, some believe that the essence of Bell's thinking lies in his gadankenexperiment in which correlations are examined for measurements at alternative instrument settings to those used, given the results at those used (counterfactuals). Inferred hypothetical results at mutually exclusive instrument settings are considered for fixed values of assumed hidden variables, whereas in [12] the hidden variables may still vary, since only measured conditioning variable values are held constant.

A major purpose of the present paper is to reconsider the Bell gedankenexperiment, and to compute a set of correlations for the three variable Bell inequality based on counterfactual reasoning. The computation makes use of nonlocal information regarding spatially separated instrument settings and measurement results. A set of correlations consistent with quantum mechanics as well as Bell's inequalities is obtained. One of the correlations is different from that usually assumed, since the counterfactual measurement must be conditional on the other measurements. The set of correlations thus corresponds to a process that is nonstationary in angle. (The same scheme may be extended to the four variable case.) The surprising feature of this correlation set that directly challenges widespread beliefs, is that it satisfies the Bell inequality in spite of having been computed through the use of nonlocal information. It follows that violation of the Bell inequality is not an infallible test of nonlocality.

A second experiment consisting of a sequence of three spin measurements on a single particle is considered as well, and the correlations contrasted with those of the Bell gedankenexperiment. The correlations among the measurements show that this process is spatially nonstationary also. However, computation of the correlations requires only local information, rather than nonlocal information, as in the previous example. In both cases, however, the mistaken assumption of *stationarity* is sufficient to cause violation of Bell's inequality.

To facilitate the discussion of the examples given in Sec. 2, the next section of the introduction will review an alternative derivation of the Bell inequality [10,11] in order to clarify the concepts used in this paper, which depart from those usually encountered in discussions of this subject.

*1.1 An arithmetic identity*

It is first shown that Bell's inequality is always satisfied by correlations of the appropriate number of lists of ± 1's. Assume the existence of three lists of numbers, *a, b,* and *b'* of length *N*, composed of elements $a_i$, $b_i$, $b'_i$, each equal to ±1. From the *i* th elements of the three lists, form the statement

$$a_i b_i - a_i b'_i = a_i (b_i - b'_i) = a_i b_i (1 - b'_i b_i). \qquad (2)$$

Sum (2) from *1* to *N*, and divide by *N*. Then take the absolute value of both sides to obtain

$$\left| \sum_{i=1}^{N} a_i b_i / N - \sum_{i=1}^{N} a_i b'_i / N \right| \le \sum_{i=1}^{N} |a_i b_i| |1 - b'_i b_i| / N = \sum_{i=1}^{N} (1 - b'_i b_i) / N . \qquad (3)$$

Note that the algebraic form of (2) and (3) depends on a factoring step as is the case in Bell's original derivation of the inequality [6]. This is because three cross-correlations are computed among the same three data sets. Note that (3) is independent of statistical assumptions, and holds even for deterministic processes. In the case of random processes such that the numerical correlation estimates in (3) approach ensemble averages as $N \to \infty$, replacing the estimates with the averages results in the usual form of Bell's inequality

$$|\langle AB \rangle - \langle AB' \rangle| + \langle BB' \rangle \ge 1 . \qquad (4)$$

It has been shown [11] that the violation of Bell inequality (4) by three correlations of the form (1) indicates that the single correlation actually observed cannot result from a stochastic process that is spatially stationary in second order correlations, i.e. one that is wide sense stationary, since triple data sets from all such processes and their resulting correlations must always satisfy (3). However, not all stochastic processes are wide sense spatially stationary, and not all random numbers need be based on a stochastic process for which readout operations commute. Nevertheless, equations (3 and (4) constrain these more exotic processes also, if the data can be arranged in three lists.

From the development above, it follows that violation of (4) by an assumed set of correlations implies that no infinite data sets exist having the cross-correlations in question. As explained in [12], the violation of (4), as well as the Clauser, Horne, Shimony, Holt (CHSH) correlational inequality in four variables [13], may occur if correlations are computed from a different number of data sets than assumed in the derivation of the relevant inequality. The assumption that the inequalities should still be satisfied under this condition, is equivalent to assuming that the underlying process is spatially stationary in angle coordinates.

## 2. Non stationary correlations for Bell's gedankenexperiment

As seen above, in the original derivation of Bell's theorem, Bell considered an EPR-Bohm experiment and posed the question: given a pair of measurements $A_i$, $B_i$, for detectors at settings $\theta_A$ and $\theta_B$, what would the corresponding value $B'_i$ at an alternate detector setting $\theta_{B'}$ have been *on that trial*, assuming the existence of fixed hidden variables to determine it? What would the correlation $\langle BB' \rangle$ be corresponding to the correlations $\langle AB \rangle$ and $\langle AB' \rangle$ for an ensemble of such trials?

Since only one pair of readings per particle pair can be experimentally obtained, the computation of $\langle BB' \rangle$ is a purely logical exercise in probability modeling. The result cannot be confirmed experimentally since the trial cannot be repeated with hidden variables held fixed (assuming that they exist) because they have not been identified. Nevertheless, the logical consequences of this gedankenexperiment lie at the heart of the Bell theorem. As indicated above, the resulting three correlations must satisfy Bell's inequality if they are based on data assumed to exist in any mathematical sense. Consequently, the three correlations cannot be represented by the usual cosine function, since the resulting violation of Bell's inequality indicates an inconsistency with the existence of any three data sets that could possibly exist.

Correlations for Bell's gedankenexperiment can be constructed using nonlocal hidden variables chosen to allow a convenient simulation of the conditional probabilities provided by quantum mechanics. Suppose the detector with output $A(\theta_A)$ is fixed in position and has returned value +1 (-1) (with probability 1/2). Then

$$p(B = \mp 1 / A = \pm 1) = \cos^2 \frac{\theta_B - \theta_A}{2}$$
$$p(B = \pm 1 / A = \pm 1) = \sin^2 \frac{\theta_B - \theta_A}{2}$$
(5)

These probabilities can be obtained from a pair of hidden variables $\lambda_1, \lambda_2$ uniformly distributed on the interval [0,1]. For the first variable $\lambda_1 \in [0, .5) \Rightarrow A = +1$, $\lambda_1 \in [.5, 1] \Rightarrow A = -1$. For the second variable (see Fig. 2),

$$\lambda_2(A = 1) \in \left[0, \cos^2 \frac{\theta_B - \theta_A}{2}\right] \Rightarrow B = -1 \; ; \; \lambda_2(A = 1) \in \left[\cos^2 \frac{\theta_B - \theta_A}{2}, 1\right] \Rightarrow B = +1. \quad (6)$$

Whereas for $A = -1$, one has

$$\lambda_2(A = -1) \in \left[0, \cos^2 \frac{\theta_B - \theta_A}{2}\right] \Rightarrow B = 1 \; ; \; \lambda_2(A = -1) \in \left[\cos^2 \frac{\theta_B - \theta_A}{2}, 1\right] \Rightarrow B = -1 \; . \quad (7)$$

Since one may determine $B$ given $A$ from hidden variables $\lambda_1$ and $\lambda_2$, one may equally well determine $B'$ in the same manner by replacing $\theta_B$ with $\theta_{B'}$. Then for angles $\theta_B$ and $\theta_{B'}$ labeled so that $\cos^2(1/2)(\theta_B - \theta_A) \geq \cos^2(1/2)(\theta_{B'} - \theta_A)$:

$$\langle BB' \rangle = \frac{1}{2} \left\{ (-1)(-1)\cos^2 \frac{\theta_{B'} - \theta_A}{2} + (-1)(+1) \left[ \cos^2 \frac{\theta_B - \theta_A}{2} - \cos^2 \frac{\theta_{B'} - \theta_A}{2} \right] + \right.$$
$$(+1)(+1) \sin^2 \frac{\theta_B - \theta_A}{2} \right\} + \frac{1}{2} \left\{ (-1)(-1)\cos^2 \frac{\theta_{B'} - \theta_A}{2} + \right. \quad (8)$$
$$\left. (-1)(+1) \left[ \cos^2 \frac{\theta_B - \theta_A}{2} - \cos^2 \frac{\theta_{B'} - \theta_A}{2} \right] + (+1)(+1)\sin^2 \frac{\theta_B - \theta_A}{2} \right\}$$

Then for $\theta_B - \theta_A$ $\theta_{B'} - \theta_A$ $\pi$

$$\langle BB' \rangle = 1 + cos(\theta_{B'} - \theta_A) - cos(\theta_B - \theta_A) \quad , \tag{9}$$

with

$$\langle AB \rangle = -cos(\theta_A - \theta_B), \quad \langle AB' \rangle = -cos(\theta_A - \theta_{B'}) \quad . \tag{10}$$

These correlations satisfy Bell's inequality (4) since

$$\left| -cos(\theta_B - \theta_A) + cos(\theta_{B'} - \theta_A) \right| + 1 + cos(\theta_{B'} - \theta_A) - cos(\theta_B - \theta_A) \quad 1 \quad . \tag{11}$$

Because $1$ may be subtracted from both sides in (11),

$$\begin{aligned} &\left| -2 \sin\tfrac{1}{2}\left((\theta_{B'} - \theta_A) + (\theta_B - \theta_A)\right) \sin\tfrac{1}{2}\left((\theta_{B'} - \theta_A) - (\theta_B - \theta_A)\right) \right| + \\ &-2 \sin\tfrac{1}{2}\left((\theta_{B'} - \theta_A) + (\theta_B - \theta_A)\right) \sin\tfrac{1}{2}\left((\theta_{B'} - \theta_A) - (\theta_B - \theta_A)\right) \quad 0 \end{aligned} \quad . \tag{12}$$

(12) yields $0$ $0$ since the arguments of the sine functions range from 0 to $\pi$.

## 3. Non stationarity due to local interaction

In this section, correlations are computed for a sequence of local measurements on a single particle. The particle begins in a mixed state such that the probability is 1/2 for measurement of positive or negative spin components in the z-direction. The initial measurement at $\theta = 0$, is followed by two additional spin measurements at detector angles $\theta_1$ and $\theta_2$. The spin value found for each measurement is determined by the particle's path through the apparatus, and this is inferred from its final position registered on a detector array. (See Fig. 3.) Spacings between detectors are assumed chosen to make this determination unambiguous. The correlations in the experiment are $\langle s(0)s(\theta_1) \rangle$, $\langle s(\theta_1)s(\theta_2) \rangle$, and $\langle s(0)s(\theta_2) \rangle$. These may be evaluated from the conditional quantum probabilities for spin:

$$p(s_{j+1} = \pm 1 / s_j = \pm 1) = \cos^2 \frac{(\theta_{j+1} - \theta_j)}{2}$$

(13)

$$p(s_{j+1} = \pm 1 / s_j = \mp 1) = \sin^2 \frac{(\theta_{j+1} - \theta_j)}{2}$$

for $j = 0, 1$. The most laborious correlation to evaluate is

$$\langle s(\theta_2) s(0) \rangle = \sum_{s_0, s_1, s_2} s_0 s_2 p(s_2, s_0) = \sum_{s_0, s_2} s_0 s_2 \sum_{s_1} p(s_2, s_1, s_0) \ , \tag{14}$$

where

$$p(s_2, s_1, s_0) = p(s_2 / s_1, s_0) p(s_1, s_0) = p(s_2 / s_1) p(s_1 / s_0) p(s_0) . \tag{15}$$

The conditional probabilities in (15) rely on non commutative operations that depend on the order in which the measurements are performed. To compute $p(s_2 = 1, s_0 = 1)$ from (13) and (14), note that

$$p(s_2 = 1 / s_1 = 1, s_0 = 1) p(s_1 = 1, s_0 = 1) = \cos^2 \frac{\theta_2 - \theta_1}{2} \cos^2 \frac{\theta_1}{2} \frac{1}{2} \ , \tag{16}$$

$$p(s_2 = 1 / s_1 = -1, s_0 = 1) p(s_1 = -1, s_0 = 1) = \sin^2 \frac{\theta_2 - \theta_1}{2} \sin^2 \frac{\theta_1}{2} \frac{1}{2} \ . \tag{17}$$

By summing over $s_1$ in $p(s_2, s_1, s_0)$, one obtains

$$p(s_2 = 1, s_0 = 1) = \cos^2 \frac{\theta_2 - \theta_1}{2} \cos^2 \frac{\theta_1}{2} \frac{1}{2} + \sin^2 \frac{\theta_2 - \theta_1}{2} \sin^2 \frac{\theta_1}{2} \frac{1}{2} \ . \tag{18}$$

Other values for probabilities $p(s_2, s_0)$ may be computed in a similar manner. One obtains

$$\langle s(\theta_2) s(0) \rangle = \cos \theta_1 \cos(\theta_2 - \theta_1) \ , \tag{19}$$

and

$$\langle s(\theta_1) s(0) \rangle = \cos \theta_1 \ ; \quad \langle s(\theta_2) s(\theta_1) \rangle = \cos(\theta_2 - \theta_1) \ . \tag{20}$$

From (4),

$$|\cos\theta_1 - \cos\theta_1 \cos(\theta_2 - \theta_1)| + \cos(\theta_2 - \theta_1) \leq 1,$$
$$|\cos\theta_1||1 - \cos(\theta_2 - \theta_1)| + \cos(\theta_2 - \theta_1) \leq 1, \quad (21)$$
$$2\sin^2\frac{\theta_2 - \theta_1}{2}\left[|\cos\theta_1| - 1\right] \leq 0.$$

The left hand side is negative or zero, so that Bell's inequality is satisfied.

## 4. Discussion

The computation of correlations treated in Sec. 2 among measurements A, B, and B' has used nonlocal information exchange. Since three data sets emerge from the analysis, Bell's inequality is satisfied, and the correlations for the process are non-stationary in angle coordinates. On the basis of a nonlocal interaction between detectors at *A* and *B*, it might be expected that a change of settings *B* to *B'* would change the data at *A*, as well as at *B*, making it impossible to construct the model shown in Sec. 2. In that case, four data streams might be thought to occur instead of three, so that the application of Bell's inequality would be inappropriate, leading to its possible violation. However, the probabilities (5), conditional on *A*, follow directly from probability theory and quantum mechanics, and the use of (5) in conventional probability theory allows one to hold A constant. This allows (5) to be simulated using a hidden variable, regardless of whether the problem involves physical nonlocality or not. Then, since one is entitled to generate the result *B*-given-*A*, one can logically generate the result *B'*-given-*A* as well by using the same method, and finally $\langle BB' \rangle$ from the ensemble average.

Cohen [14] has shown that in the case of multiple continuous variables, an infinite number of joint probability densities can in general be found from given marginal densities. It is thus not surprising that non-uniqueness has been encountered in the cases of discrete variables in Secs. 2 and 3. In these examples, two of the three correlations are given by the cosine function (with sign reversed between the examples), while the form of the third correlation depends on different physical specifications of the two situations. It is thus seen that two of three correlations do not determine the third. Each set of correlations exhibits an overall rotational symmetry, but the third correlations from each set are different.

It should be remarked that inequalities for probabilities of outcomes of the Bell experiments have been derived [15], and that their relation to the corresponding correlational Bell's inequalities such as that of CHSH [13] has never been spelled out. The author has found it possible to derive the Clauser-Horne (CH) inequality [15] directly from the correlational form by assuming that the probabilities have the following symmetry: $P_{++}(\theta_A, \theta_B) = P_{--}(\theta_A, \theta_B)$, and $P_{+-}(\theta_A, \theta_B) = P_{-+}(\theta_A, \theta_B)$. The CH inequality is thus less general than the correlational form, which holds even for deterministic data-list cross-correlations. The key result is that when consistent sets of spatially nonstationary correlations satisfying the correlational inequality are found, the probabilities corresponding to them satisfy the equivalent CH inequality, provided that the symmetry conditions are satisfied. For the sets of correlations of interest in quantum mechanical experiments (which satisfy the symmetry conditions), violation of the probability inequality implies violation of the correlational inequality, which then implies that no infinite data sets exist with the required cross-correlations. The correlational and probability forms of Bell inequality are not logically independent. It is intended to treat this topic in more detail in a future submission.

## 5. Summary and conclusion

Several recent results concerning the interpretation of Bell's inequalities may now be summarized. The Bell's inequality for correlations was historically derived on the assumption of a stochastic process characterized as wide-sense-stationary in the angle coordinate. In fact, it has been shown to hold generally, without such a restrictive assumption, for cross-correlations of any three data sets consisting of $\pm 1$'s. Since satisfaction of the inequality is independent of the size of the data sets, it holds for correlations in the limit of infinite experimental data. (Equivalent statements hold for four data sets.) The data may be completely deterministic or randomly generated by any kind of process whatsoever. Proposed sets of correlations violating the inequality therefore cannot arise from cross-correlations of any three infinite data sets that can mathematically exist.

Historically, correlations have been computed from a number of experimental data sets that is not the same as the number assumed in the derivation of the Bell inequality. These correlations are therefore not logically required to satisfy the inequality. The assumption that they should, is tantamount to assuming that they are based on a wide-sense-stationary stochastic process, as was indicated by Bell. But that assumption is inconsistent with the non commutativity of measurements of spins or polarizations. The computation of correlations of spin or polarization measurements consistent with the Bell inequality derivation leads to nonstationary correlations. This is true both for real experimental measurements, and for theoretically generated counterfactual data. In the examples shown, nonstationary correlations satisfy the Bell inequality both when nonlocal information is used in the computation, and when it is not. However, if a single function is assumed to characterize all the measurement correlations of these nonstationary processes, Bell's inequality is violated at certain detector settings. The foregoing results strongly challenge the common belief that violation of Bell's inequality indicates nonlocality.

Analogous statements hold for inequalities based on probabilities. The Clauser Horne (CH) inequality can be derived from the CHSH correlational form, given symmetries in the probabilities that characterize physical problems of interest. Nonstationary correlations that satisfy the CHSH inequality correspond to probabilities that satisfy the CH inequality.

These foregoing considerations leave open the historic question of whether a local hidden variable model exists for the Bell cosine correlation. The above discussion implies that any proposed candidate model for the correlation must have a nonstationary character to be consistent with the non commutation requirement of quantum mechanics.

**ACKNOWLEDGEMENT**

I am indebted to Michael J. Steiner for several fruitful discussions on the subject of the manuscript,

and for raising the issue considered in Sec. 3.

FIGURE CAPTIONS

FIG. 1   Particles in singlet state are emitted in opposite directions and detected by Stern-Gerlach apparatus with orientations $\theta_A$ and $\theta_B$ with alternate settings $\theta_{A'}$ and $\theta_{B'}$.

FIG. 2   Given A = -*1*, a uniformly distributed random variable $\lambda_2$ implies $B(\theta_B) = \pm 1$ as $\lambda_2$ is less or greater than $cos^2(\theta_B/2)$. If A = +1, the signs of $B(\theta_B)$ are correspondingly reversed. To employ the rule, an agent at B would need to know the angular setting at A and its measurement outcomes.

FIG. 3   Schematic representation of apparatus consisting of three Stern-Gerlach magnets for three sequential spin measurements.

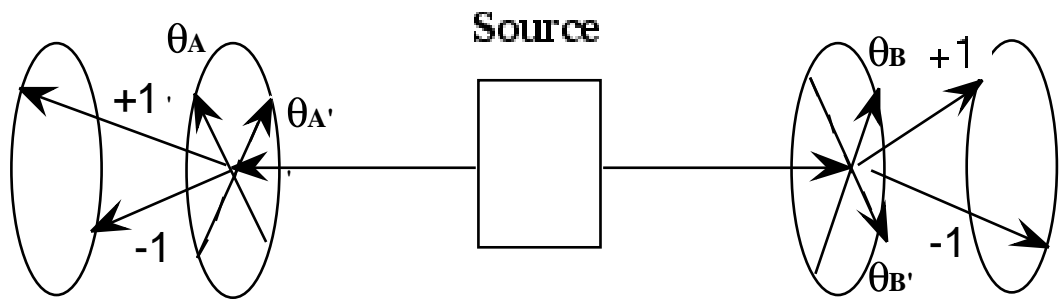

Fig. 1

# Diagram for Hidden Variable Construction

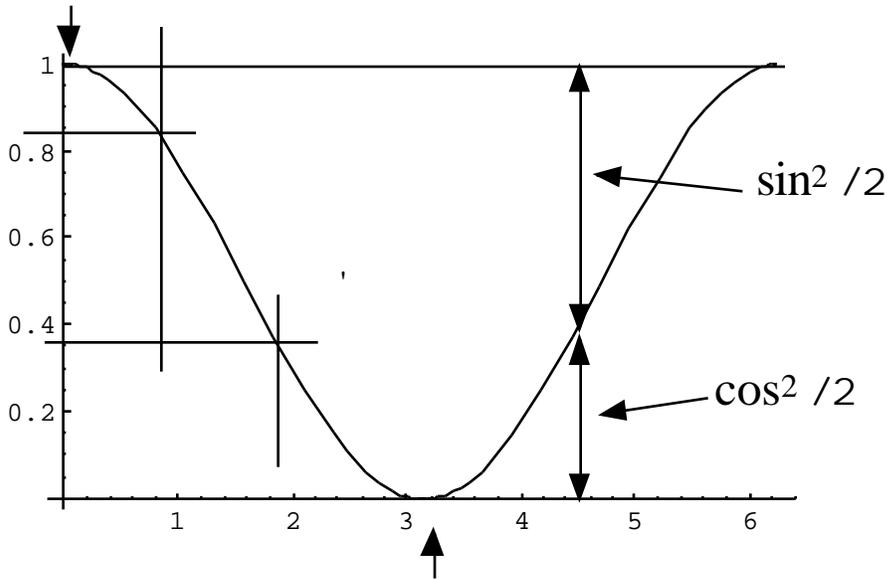

Fig. 2

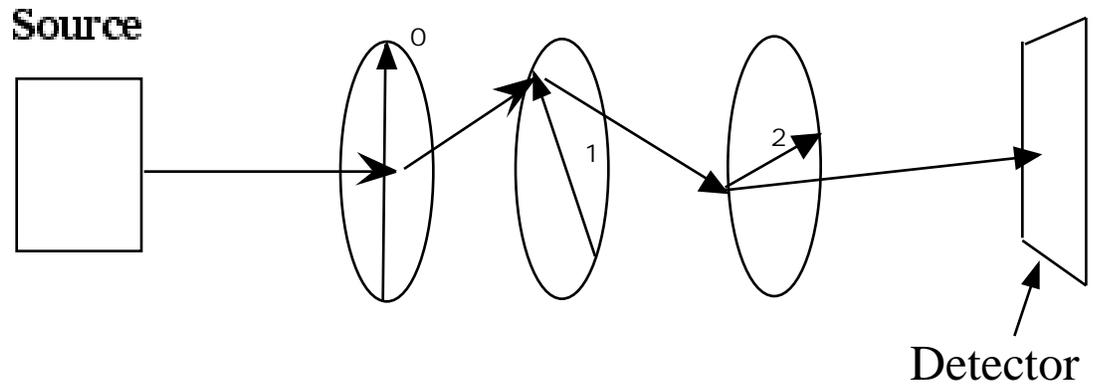

Fig. 3